\documentclass[12pt,preprint]{aastex}
\usepackage{graphicx,lscape}
\def\ea{\it et al.}

\def\ros{{\sl ROSAT}}
\def\cha{{\sl Chandra}}
\def\xmm{XMM-{\sl{Newton}}}
\def\psr{{PSR~B0628-28}}
\def\ergsec{\hbox{erg s$^{-1}$ }}
\def\ergcm{\hbox{erg cm$^{-2}$ s$^{-1}$ }}

\def\lx{$L_{\mbox{x}}$}

\shorttitle{X-ray emission from \psr}
\shortauthors{Tepedelenlio\v{g}lu \& \"{O}gelman}

\begin{document}
\title{\cha\ and \xmm\ observations of the exceptional pulsar B0628-28}

\author{E. Tepedelenlio\v{g}lu\altaffilmark{1}
\email{emre@cow.physics.wisc.edu}
\and
H. \"{O}gelman\altaffilmark{1,2}}
\email{ogelman@cow.physics.wisc.edu}
\altaffiltext{1}{Department of Physics, University of
Wisconsin-Madison, 1150 University Ave., Madison, WI 53703, USA}
\altaffiltext{2}{Faculty of Engineering and Natural Sciences, Sabanci
University, Orhanli Tuzla, Istanbul 34956, Turkey}

\begin{abstract} 
\psr\ is a radio pulsar which was first detected in the X-ray band by
\ros\ and then later observed with \cha\ and \xmm. The \cha\
observation yielded an X-ray luminosity two orders of magnitude higher
than what is expected for spin-powered pulsars, also there were no
pulsations detected. The \xmm\ observation, however, reveals
pulsations at the expected radio period, $P=1.244$ s. The
simultaneously analyzed spectra also give a luminosity (in cgs)
$\log{L_{\mbox{x}}}=30.34$, which is $\sim 350$ times greater than
what would be expected from the correlation between
$L_{\mbox{x}}$-$\dot{E}$. 
\end{abstract}


\section{Introduction}\label{intro}
\psr\ is an old pulsar that has a characteristic age
($\tau=P/2\dot{P}$) of 2.8 Myr. Since older pulsars have radiated away
their initial heat content and have relatively low rotational energy
loss ($\dot{E}=I\Omega\dot{\Omega}$), one would expect to detect X-ray
radiation coming either from the reheating of the surface
\citep[e.g. see][]{tsuruta,shibazaki} or by heated polar caps
\citep{halpern}. In young pulsars, however, thermal radiation from the
neutron star surface is dominated by the nonthermal component from the
neutron star magnetosphere, whose spectrum can be described primarily
by a power-law model. We, also, know from X-ray observations that the
pulsar's nonthermal luminosity (\lx) shows a correlation with the
spin-down power $\dot{E}$. For instance, \citet{possenti} found a best
fit based on 37 pulsars. Although the variance is large, pulsars
follow a general trend that can be formulated as
$\log{L_{\mbox{x}}^{2-10}}=1.34\log{\dot{E}}-15.34$. This set included
only three old ($\tau > 10^{6}$~years) pulsars. The reason for having
only three is two fold. The first one is that older pulsars are less
active and relatively cooler, hence most of them were not detected
until \cha\ and \xmm. The detected ones have low countrates,
consequently statistically poor spectra. This makes it challenging to
distinguish between spectral models. The other reason is that it is
expected that surface emission from old neutron stars will be
dominant, resulting in spectra to be more like a blackbody. Hence, the
X-ray radiation from these pulsars are not expected to be
characterized by a power-law model. However, recent observation of
PSR~B0943+10 \citep{zhang}, PSR~B0823+26, PSR~B0950+08, PSR~J2043+2740
with \xmm\ \citep{becker} and of PSR~B2224+65 with \cha\
\citep{zavlin} has unfolded a new perspective in understanding old
neutron stars. Not only many previously undetected sources were
detected but also it has been possible to distinguish between
different spectral models for some of these old pulsars. These old
pulsars have spectra harder than expected and is best described by a
power-law and/or a polar cap (black-body) model, as opposed to a
black-body model representing the thermal emission from the whole
neutron star surface. The integrated model flux of these sources
convert to unusually high X-ray efficiencies ($L_{\mbox{x}}/\dot{E}$).

\psr\ is the longest period radio pulsar detected in X-rays. However,
physical parameters inferred from radio observations are not in any
means extreme. The \cha\ data yielded a luminosity two orders of
magnitude greater than what is expected for a spin-powered pulsar
\citep{ogelman}. The varying efficiency of converting spin-power in to
X-ray luminosity in pulsars can be explained by geometrical
effects. But, non of these can amount to such a great excess flux in
the X-ray band. In this paper we try to address the reasons for this
extreme luminosity. We describe the \xmm\ and \cha\ data in
$\S$~\ref{obs} and present the result of our timing and spectral
analysis in $\S$~\ref{res}. We discuss the implications of the
analysis results in $\S$~\ref{dis}.

\section{\cha\ and \xmm\ observations of \psr}\label{obs}
\subsection{\cha} 
\psr\ was observed twice, first on 2001 November 04 and again on 2002
March 25 for 2000 s and 17000 s, respectively. For both observations
photons were collected using the Advanced CCD Imaging Spectrometer
(ACIS). Data were collected in the nominal timing mode, with 1.141 s
exposures between CCD readouts. We reprocessed the Level 1
event data to correct the detrimental effects of charge transfer
efficiency.  The imaging, timing and spectral analysis presented here
were done only on the 17 ks observation, the 2.0 ks observation was
disregarded due to an interruption by a large solar storm. The
background countrate during this solar storm increased by a factor of
$\sim$20. The net source countrate during the first observation period
was $0.011\pm 0.001$ counts per second (cps) where the error is in the
68\% confidence range. Since by taking into account the 2.0 ks
observation we gain only $\sim$27 counts, which is not enough to
improve our statistics significantly, we preferred to disregard these
counts, which potentially can be misleading.

The measured PSF of \psr\ is consistent with the ACIS point-source
response, hence the ACIS image reveals a point-like X-ray source at the
pulsar position. The Chandra position of \psr\ is
$\alpha=06^{\mbox{h}}30^{\mbox{m}}49\farcs43$,
$\delta=-28^{\circ}34^{\prime}43\farcs60$ (J2000.0), which considering
0\farcs5 rms error and the $\sim$0\farcs6 ~absolute astrometric
accuracy of \cha, is in good agreement with the radio position
$\alpha$=06$^{\mbox{h}}$30$^{\mbox{m}}$49\farcs53, $\delta=-28^{\circ}34
\arcmin43\farcs60$ (J2000.0), which was taken from the ATNF pulsar
catalogue{\footnote{http://www.atnf.csiro.au/research/pulsar/psrcat}}.

\subsection{\xmm}

\psr\ was observed with \xmm\ on 2004 February 28 for a total ontime
of 48 ks. MOS1/2 were both operated in imaging (PrimeFullWindow) mode
and the medium filter was used. During the EPIC-pn exposure the thin
filter was used and the detector was operated in imaging
(PrimeLargeWindow) mode for 47 ks. The temporal resolution achieved
with this choice of science modes were 2.6 s and 43 ms for MOS1/2 and
pn, respectively.

The background of EPIC camera is known to be effected by soft proton
flares. In order to screen for times of high background we inspected
the light curves of MOS1/2 and pn data separately at energies above 10
keV. Both sets of data were effected by high background. The light
curves were formed with 100 s bins and after inspection we rejected
bins with countrate greater than 0.4 cps. The removal of high
background time intervals from the data leaves us with effective
exposure times of 42.5 ks and 33.3 ks for MOS1/2 and pn,
respectively. The pulsar is clearly detected in the EPIC image at
$\alpha=06^{\mbox{h}}30^{\mbox{m}}49\farcs48$, $\delta=-28^{\circ}
34^{\prime}43\farcs10$ (J2000.0) which differs from the radio position
by only 0\farcs8 well within the 2\arcsec-3\arcsec\ uncertainty of the
EPIC absolute astrometry. The shape of the radial profile of the
source is also consistent with that expected for a point-like source.

\section{Results}\label{res}
\subsection{Timing}
For searching pulsations from \psr\ \cha\ ACIS and \xmm\ EPIC-MOS data
were not suitable due to their limited temporal resolution. The
sampling frequency in both cases set by the detector read out rate give
a Nyquist frequency greater than the pulsar frequency (see
Table~\ref{tab1}). Thus we used EPIC-pn data that has 43 ms timing
resolution.

\begin{figure}
\begin{center}
\includegraphics[angle=270,width=16cm]{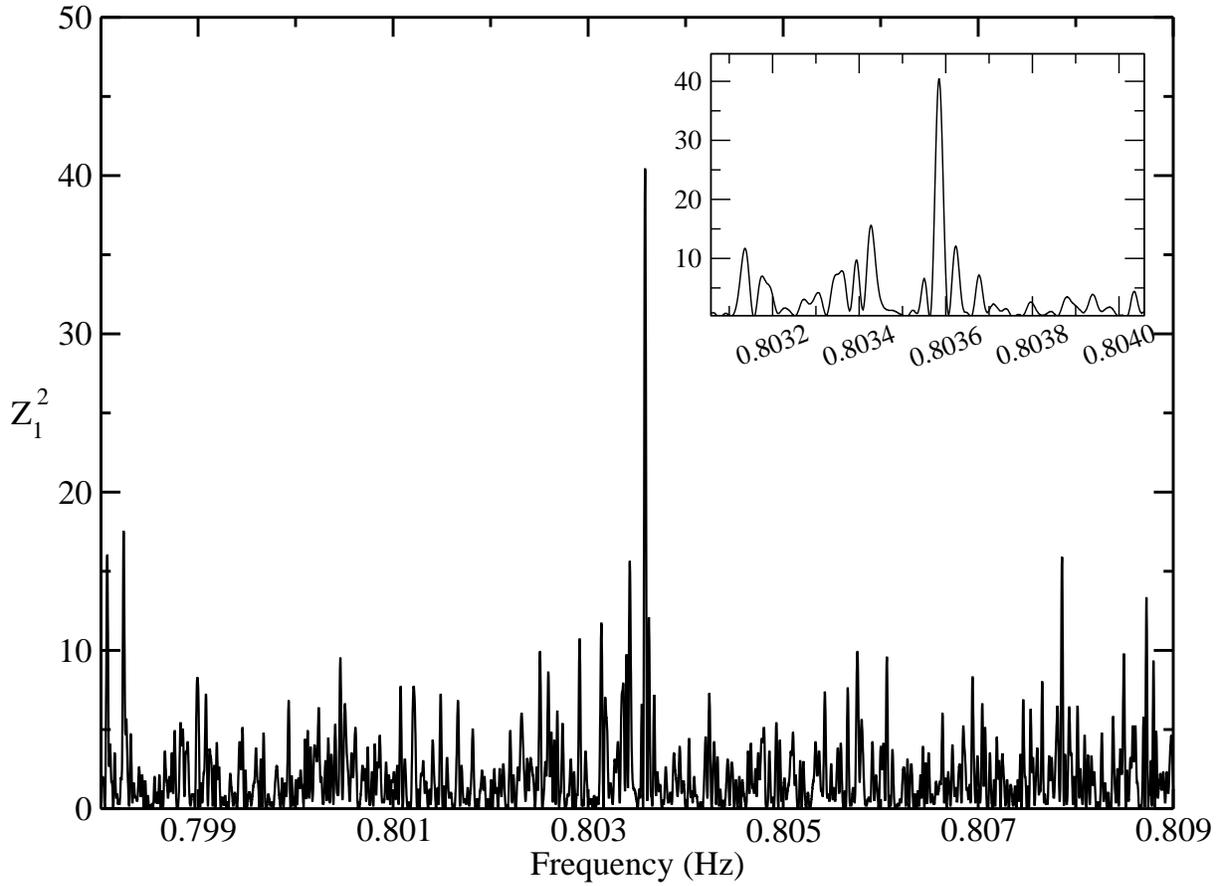}
\end{center}
\caption{\small{The power spectrum of the timing data around the
expected frequency. The inset is the expanded view $\pm 500$ $\mu$Hz
around the peak, which is located at $f=0.80358551$ Hz.}}
\label{fig1}
\end{figure}

\begin{figure}
\begin{center}
\includegraphics[angle=270,width=16cm]{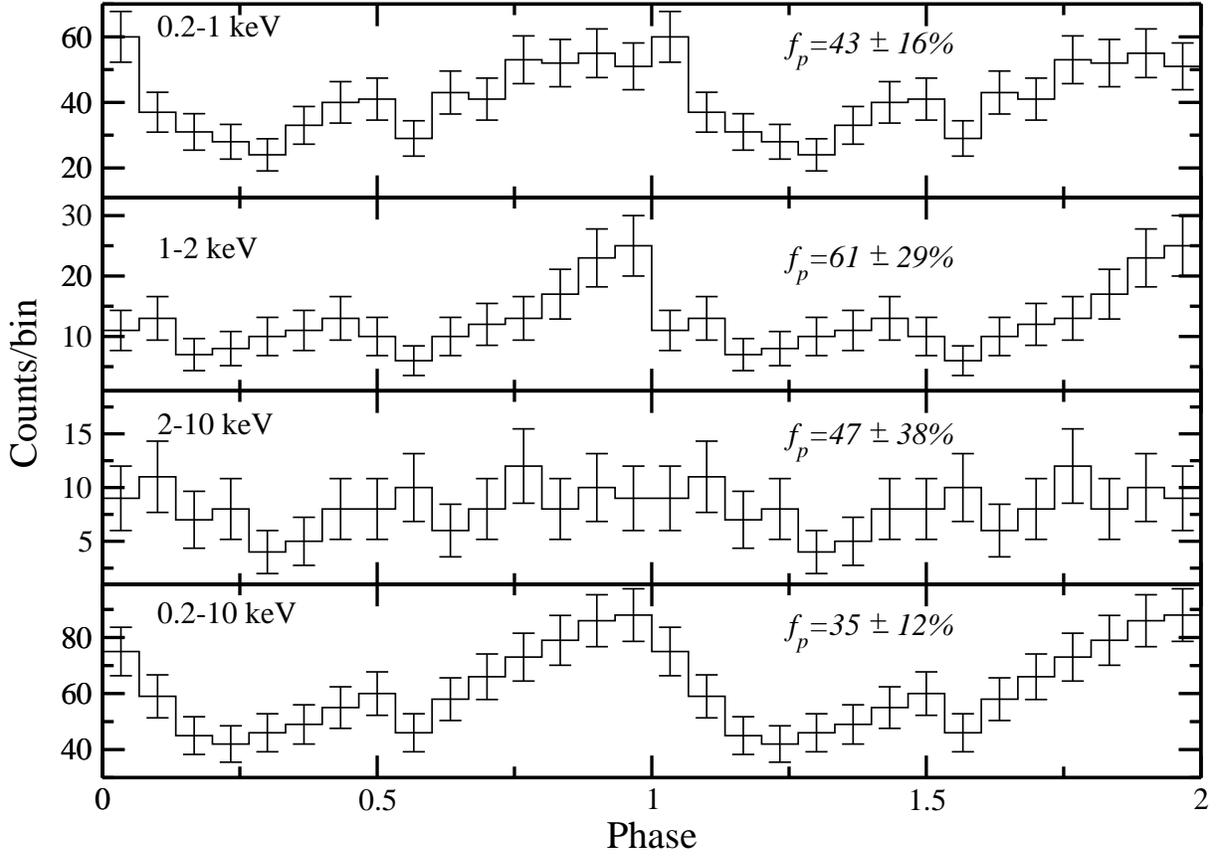}
\end{center}
\caption{\small{X-ray light curves of \psr\ extracted from the EPIC-pn
data in four energy bands, with the values of the intrinsic pulsed
fraction $f_{p}$ and its 1$\sigma$ errors. Two phase cycles are shown
for clarity. For the expression used for the pulse fraction see
text.}}
\label{fig2}
\end{figure}

We extracted source plus background photons from a 30\arcsec\ radius
circle centered at the pulsar position, which encircles about 85\% of
all detected source counts. The extraction region contained 1047
counts of which 16\% is background. The photon arrival times were
solar system barycenter corrected. The pulsar's spin parameters
(Table~\ref{tab1}) are well known from radio observations and can be
extrapolated to the mean epoch of the \xmm\ observation:
MJD=53063.339. Around this predicted pulsar frequency, we then
generated a periodogram using the $Z_{1}^{2}$-statistic. The X-ray
periodogram is shown in Figure~\ref{fig1}. We found a peak at
$f=0.80358444\pm 0.00000112$ Hz which is consistent with the
extrapolated pulsar frequency $f=0.80358551$ Hz. The $Z_{1}^{2}$ for
this peak is 40.4, which has a probability of chance occurrence of
1.69$\times 10^{-9}$. The pulse profile of \psr\ over the whole energy
band (bottom panel in Figure~\ref{fig2}) is broad and single-peaked
with a pulse fraction of $f_{p}=35\pm 12$\%. Here we defined the pulse
fraction as
$(C_{\mbox{max}}-C_{\mbox{min}})/(C_{\mbox{max}}+C_{\mbox{min}})$,
where $C_{\mbox{max}}$ and $C_{\mbox{min}}$ are the maximum and
minimum counts per bin, respectively.

We also looked at the pulse profile of the pulsar at different energy
bands (upper panels of Figure~\ref{fig2}). Most of the counts are in
the soft band, 67\% and 87\% of all counts are in the 0.2-1 and 0.2-2
keV energy bands, respectively. Pulsed fraction in each selected
energy band seems to be consistent with each other within the
1$\sigma$ errors associated. The pulse shape does not seem to be
energy dependent which would suggest that all pulsed X-rays are coming
from the same region.

\subsection{Spectral}\label{spec}

The pulsar's energy spectrum was extracted from the MOS1/2 data by
selecting all events detected in a circle of radius 50\arcsec\
centered on the pulsar position. This region includes 90\% of all
event from the pulsar. Due to a source located close to the pulsar we
extracted the background spectrum from a nearby circular region with
radius 87\arcsec. For the EPIC-pn data we extracted the spectrum from
a circle centered on the pulsar with radius 30\arcsec\ (includes 80\%
of source counts). \psr\ was being located very close to the chip
boundary precluded the extraction of the background spectrum from an
annular region around the pulsar. Hence, we used an off-source
circular region with radius 66\arcsec.

To extract the spectrum from the \cha\ data we used a circular region
centered on the pulsar position with 2\arcsec\ radius. This region
contains 95\% of all source counts. The background spectrum was
extracted from an annulus of radii 3\arcsec$<r<$50\arcsec. 

In total, the extracted spectra include 780 EPIC-pn source counts and
754 EPIC-MOS1/2 source counts. Both spectra were binned so that each
bin contained minimum 25 counts per bin. \cha\ data had a total of 184
source counts. The extracted photons were binned and regrouped such
that each fitted spectral bin contained minimum of 20 counts. All
three extracted spectra were then simultaneously fitted with model
spectra.

\begin{figure}
\begin{center}
\includegraphics[angle=270,width=16cm]{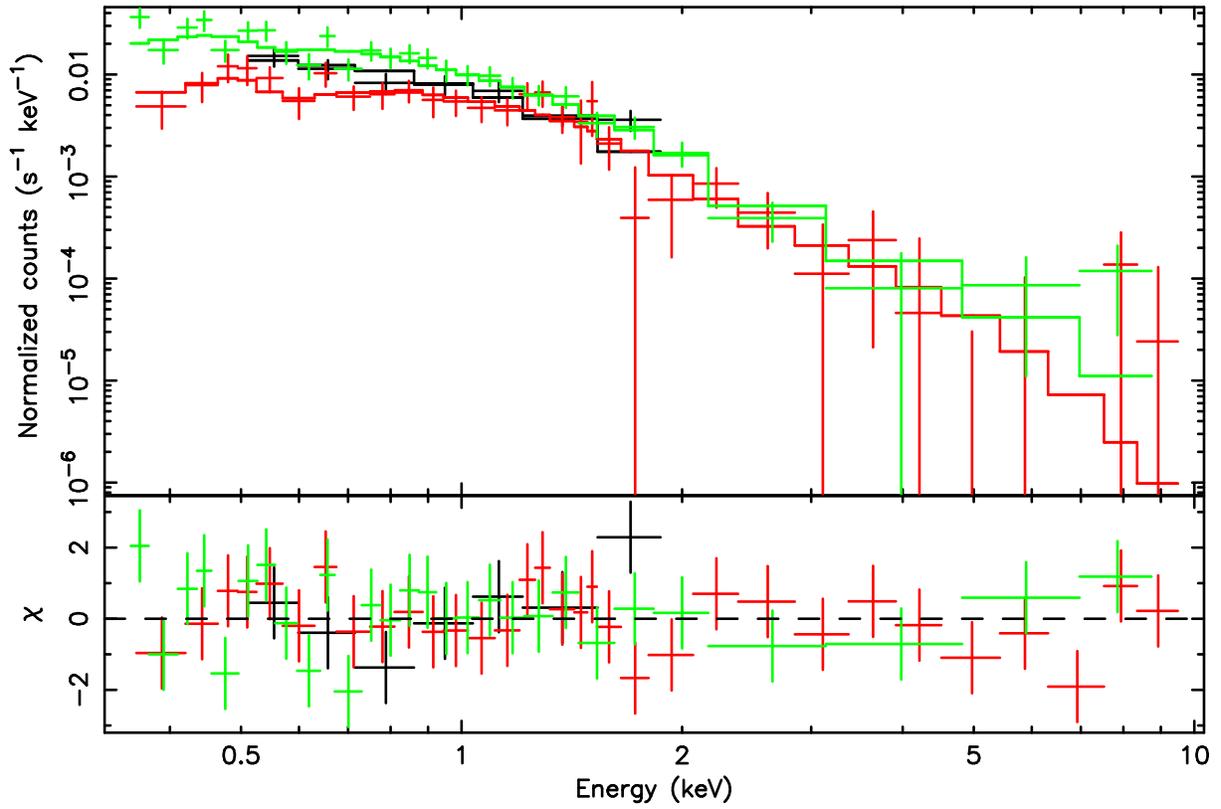}
\end{center}
\caption{\small{Energy spectrum of \psr\ as observed with pn (upper
spectrum), MOS1/2 (lower spectrum) and ACIS-S (middle spectrum). The
lines are the best fit power-law model. The lower panel is fit
residuals.}}
\label{fig3}
\end{figure}

Among the single-component spectral models, an absorbed power-law
model gave the statistically best representation ($\chi^{2}=56.2$ for
62 degrees of freedom (dof)) of the observed spectrum. A single
black-body ($\chi^{2}=93$ for 62 dof) did not give a statistically
acceptable fit. This fit when absorbing column is left to vary also
yields a very low $N_{\mbox{H}}$. The best fit power-law spectrum and
residuals are shown in Figure~\ref{fig3}. 

The power-law model yields a column density of
$N_{\mbox{H}}=1.38^{+0.37}_{-0.23}\times10^{21}$ cm$^{-2}$, a photon
index $\Gamma=3.20^{+0.26}_{-0.23}$, and a normalization of
$1.73^{+0.26}_{-0.22}\times10^{-5}$ photons cm$^{-2}$ s$^{-1}$
keV$^{-1}$ at $E=1$ keV. The errors are the upper and lower bounds of
the 1$\sigma$ confidence range. The normalization converts to an
unabsorbed energy flux of $f^{2-10}_{\mbox{x}}
=8.61^{+2.15}_{-0.33}\times10^{-15}$ \ergcm in the 2-10 keV
band. Given the distance of $d=1.45$ kpc this yields an X-ray
luminosity of $L^{2-10}_{\mbox{x}}
=2.17^{+0.56}_{-0.07}\times10^{30}$ \ergsec. This luminosity implies a
rotational energy to X-ray energy conversion factor
$L_{\mbox{x}}/\dot{E}= 0.015$ within 2-10 keV band.

It is natural to assume that, in addition to the magnetospheric
emission, thermal emission from polar caps or from the surface due to
reheating of the superfluid interior contributes to the observed X-ray
flux. In order to explore this possibility and how it represents the
data we used a two component model, thermal and magnetospheric, to fit
the spectra. As a first approach we let every parameter vary. The fit
gives a blackbody temperature of $T=3.28_{-0.62}^{+1.31}\times 10^{6}$
K and effective radius $R=59_{-46}^{+65}$ m. The power-law photon
index does not change significantly and has a value of
$\Gamma=2.98_{-0.65}^{+0.91}$. The bolometric luminosity of the
thermal component is $L_{\mbox{bol}}=2.87\times 10^{30}$ \ergsec
whereas the total luminosity in the 2-10 keV is $L^{2-10}_{\mbox{x}}
=1.67^{+0.91}_{-0.62}\times10^{30}$ \ergsec. The nonthermal X-ray
luminosity in the 2-10 keV band converts to an X-ray efficiency of
$L_{\mbox{x}}/\dot{E}= 0.01$. The hydrogen column density is not well
bound $N_{\mbox{H}}=0.62^{+0.98}_{-0.62}\times10^{21}$ cm$^{-2}$ but
still is consistent with the estimate obtained from the single
power-law fit. The quality of the composite model ($\chi^{2}=54.5$ for
62 dof) is slightly better than that obtained from single power-law
model. However, both fits are statistically acceptable and addition of
a new model will always tend to decrease the $\chi^{2}$.

\section{Discussion}\label{dis}

With the observations of new sensitive X-ray telescopes like \xmm\ and
\cha\ we are uncovering a class of new sources; X-ray luminous
old-radio pulsars. Up to now there are only seven such sources that
have been detected. Namely, these are PSRs B2224+65, J2043+2740,
B1929+10, B0823+26, B0950+08, B0943+10 and B0628-28
\citep[see][]{zavlin,becker,zhang,ogelman}. Of these six sources only
two of them (PSRs B0950+08, B0628-28) have high quality spectra so
that one can distinguish between spectral models (e.g. thermal
vs. nonthermal). \citet{becker} suggested that the three pulsars
observed with \xmm\ (PSRs B0950+08, 0823+26, J2043+2740) all have
spectra described better by a single power-law model, indicating
nonthermal emission is dominant. For these pulsars, however, there are
alternate representations that do give statistically and physically
acceptable results \citep{zavlin}. For example, \citet{zavlin} argues
that the soft part of the spectrum of PSR B0950+08 can be interpreted
as radiation from heated polar caps on the neutron star surface
covered with a hydrogen atmosphere. For PSR B2224+65 \citet{zavlin}
arrives at the conclusion, from the analysis of the \cha\ ACIS
observation, that the pulsar definitely shows non-thermal
emission. All these pulsars have X-ray luminosities greater than what
is predicted from the correlation between
$L_{\mbox{x}}^{2-10}$-$\dot{E}$ \citep{possenti}. For example, in the
case of PSR B0628-28 and PSR B2224+65 the luminosities inferred from
spectral fits are greater then the so-called ``maximum efficiency
line'' derived by \citet{possenti} such that all pulsar lie below this
line in the $\log{L_{\mbox{x}}}$-$\log{\dot{E}}$ plane. \psr\ has been
known to be an exceptional emitter and with the detection of
pulsations from this source at the radio frequency has left no doubt
that it is the X-ray counterpart of the pulsar. These luminous old
pulsars (in particular \psr) seem not to follow the trend their
possible progenitors do. Which suggests that pulsars become more X-ray
efficient as they grow older, given that $\tau$ is inversely
proportional to $\dot{E}$. From multi-wavelength observations
\citet{zharikov} have arrived at the same conclusion. However,
\citet{zavlin} suggest that the most plausible reason for the
observed high nonthermal X-ray efficiencies associated with the old
pulsars is the geometrical effects. We discuss these below.

\begin{figure}
\begin{center}
\includegraphics[angle=0,width=16cm]{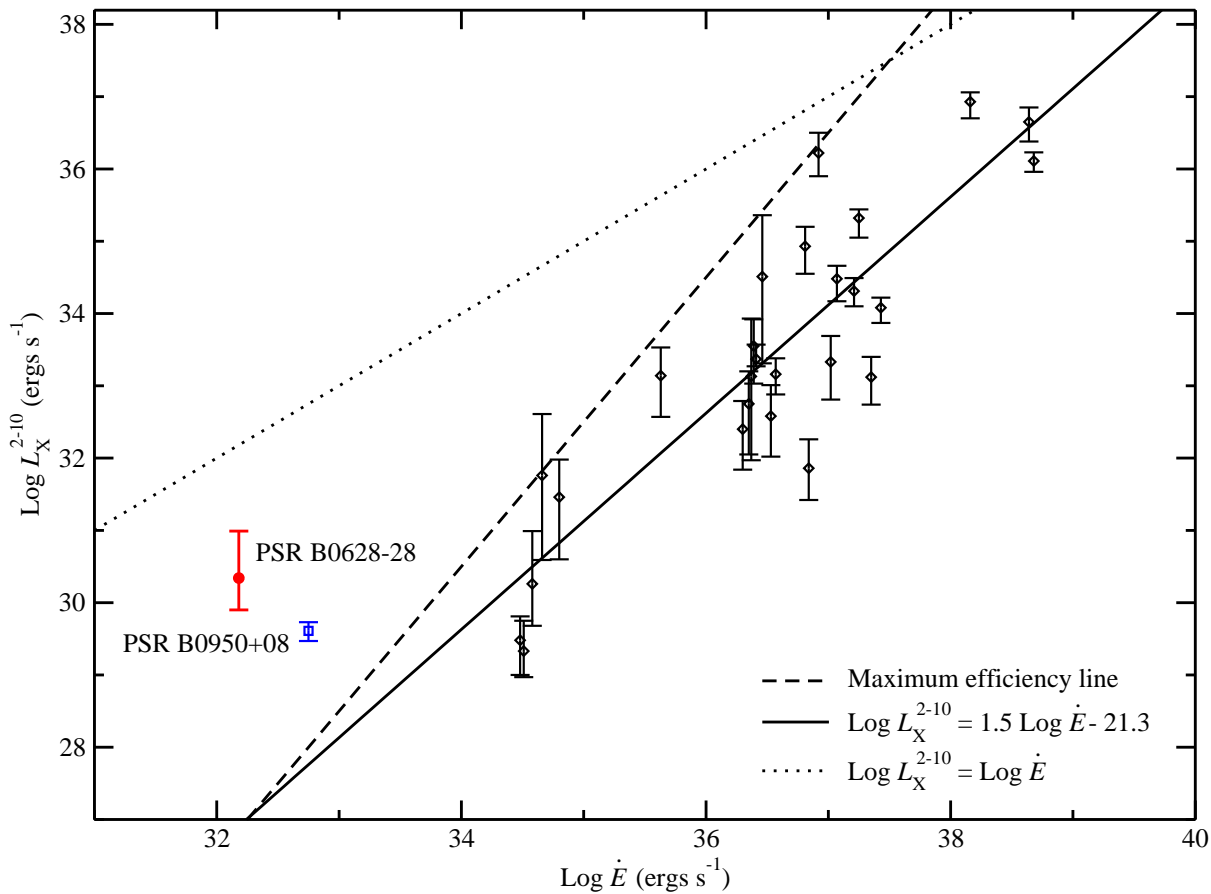}
\end{center}
\caption{\small{The X-ray luminosity in the band 2-10 keV versus the
spin-down power for the 26 sources together with two old pulsars. The
value for the X-ray luminosity of PSR B0950+08 is taken from
\citet{becker} and the rest are taken from \citet{possenti}. The solid
line is the best fit to the set of 26 pulsars. Other two lines are
labeled in the legend.}}
\label{fig4}
\end{figure}

As mentioned in $\S$~\ref{intro}, \citet{possenti} used data from 39
X-ray emitting pulsars to find a best fit that describes the
correlation between $\log{L_{\mbox{x}}}$-$\log{\dot{E}}$. These 39 sources
were divided in to subcategories as; {\it{millisecond pulsars}}
($P\la$10 ms), {\it{Crab-like}} ($\tau\sim 10^{4}$ years),
{\it{Vela-like}} ($\tau\sim 10^{4}$-$10^{5}$ years),
{\it{Geminga-like}} ($\tau\ga 10^{5}$ years, where substantial amount
of the X-ray flux comes from the internal cooling) and finally
{\it{Old-pulsars}} ($\tau\ga 10^{6}$ years). In Figure~\ref{fig4} we
show a similar plot. In this plot we only include Crab, Vela and
Geminga-like pulsars. We did not include millisecond pulsars because
they are re-cycled and do not represent the naturally aging pulsars
with the same $\dot{E}$. Also when converting countrates in to
luminosities, \citet{possenti} adopted a power-law spectrum with
$\Gamma=2$ for all millisecond pulsars. Hence, the luminosities do not
come from spectral analysis and rather from assumptions based on other
pulsars. Also we excluded the three Old-pulsars because their fluxes
were obtained by scaling their \ros\ countrates to that of PSR
B1929+10.

This newly formed set of pulsars represent the evolution of pulsars on
the $\log{L_{\mbox{x}}}$-$\log{\dot{E}}$ plane. Where older pulsars
are on the bottom left (low $\dot{E}$) and young Crab-like pulsars are
on the top right (high $\dot{E}$). Using only these three subclasses
(26 sources) we performed a linear fit of the form
$\log{L_{\mbox{x}}}=a\log{\dot{E}}+b$ (see Figure~\ref{fig4}). Due to
exclusion of the stated pulsars our fit yields a line with a steeper
slope ($a\sim 1.5$). We also identified the line of maximum efficiency
as the line for which every pulsar lies underneath. We then overlayed
only two of the seven old pulsars, with well known spectra. We
should note that, however, with their luminosities derived from the
power-law fits, PSR B0943+10 \citep{zhang} and PSR B2224+65
\citep{zavlin}, have efficiencies exceeding the maximum efficiency.

From Figure~\ref{fig4} it is apparent that old pulsars have very high
nonthermal X-ray efficiencies. This is contrary to what
\citet{possenti} has found and to what has been found when millisecond
pulsars were excluded. The observed luminosities for younger pulsars
also show large deviations, from this dependence. But in their case
the deviation is symmetric around the trend line. The scatter that
young pulsars exhibit could be due to uncertainties in pulsar
distances, spread in the orientations of magnetic and rotational axes
versus the line of sight. For example, seeing a certain fraction of
the beam would result in lower inferred X-ray efficiency and vice
versa.

Although there should be a thermal contribution to the overall
luminosity, this effect will not change the nonthermal luminosity
significantly. For instance, when we subtract the thermal luminosity,
obtained for \psr\ (see $\S$~\ref{spec}), from the total, the
nonthermal luminosity only changes by a factor of 0.5. The resulting
luminosity is still too big and gives a nonthermal X-ray efficiency
2-3 orders of magnitudes greater than the maximum efficiency.

The quality of the observed spectrum for these old pulsars are in
general not high enough to distinguish between a thermal and a
non-thermal model, or a combination of both. The current set of old
pulsars all either show a non-thermal emission or the statistics is
not high enough to rule it out. For example, \citet{becker} suggests
that the X-ray emission from old, nonrecycled rotation powered pulsars
is dominated by non-thermal radiation. When the data is fit by a
power-law model, which in all cases is statistically acceptable, the
obtained power-law indices are on average larger than the younger
rotation powered pulsars. The mean of the photon indices of the 26
pulsars (Figure~\ref{fig4}) is $\sim$1.9 as opposed to $\sim$2.5 for
the old pulsars.  The power-law fits suggests that the spectra of
these old pulsars being steeper should result in lower luminosities in
the 2-10 keV band. However, we observe an inverse effect where not
only the spectra are steeper but also they have higher
luminosities. Further X-ray observations of old radio pulsars should
help us understand these puzzling features.

\newpage
\begin{deluxetable}{ll}
\tablewidth{0pc}
\tablecaption{Properties of \psr\ derived from radio observations
\label{tab1}}
\tablehead{Parameter & Value}
\startdata
Frequency (Hz)...............................& 0.80358811986\\
Frequency derivative (Hz s$^{-1}$)........&-4.59962$\times 10^{-15}$\\
Epoch (MJD).................................&46603.0\\
Spin-down age ($10^6$ yr)..................& 2.77\\
Spin-down energy ($10^{32}$ ergs s$^{-1}$)...& 1.5\\
Inferred Magnetic Field ($10^{12}$ G)...& 3.0\\
Dispersion Measure (pc cm$^{-3}$).......& 34.5\\
Distance (kpc)................................& 1.45\\
\enddata
\tablecomments{Distance is inferred from dispersion-measure 
\citep{cordes}.}
\end{deluxetable}

\newpage
\newpage
\newpage
\newpage

\end{document}